# Observation of oriented Landau levels in Berry dipole semimetals

Qingyang Mo[1], Riyi Zheng[3], Cuicui Lu[1,2§], Xueqin Huang[3‡], Zhengyou Liu[4], Shuang Zhang[1,5,6†]

[1] New Cornerstone Science Laboratory, Department of Physics, University of Hong Kong; 999077, Hong Kong, China

[2] Key Laboratory of Advanced Optoelectronic Quantum Architecture and Measurements of Ministry of Education, Beijing Key Laboratory of Nanophotonics and Ultrafine Optoelectronic Systems, School of Physics, Beijing Institute of Technology, Beijing 100081, China

[3] School of Physics and Optoelectronics, South China University of Technology, Guangzhou, Guangdong 510640, China

[4] Key Laboratory of Artificial Micro- and Nanostructures of Ministry of Education and School of Physics and Technology, Wuhan University, Wuhan 430072, China

[5] Department of Electrical & Electronic Engineering, University of Hong Kong; 999077, Hong Kong, China

[6] Quantum Science Center of Guangdong-Hong Kong-Macao Great Bay Area, 3 Binlang Road, Shenzhen, China

§cuicuilu@bit.edu.cn
‡phxqhuang@scut.edu.cn
†shuzhang@hku.hk

# Abstract

Band crossing points, such as Weyl and Dirac points, play a crucial role in the topological classification of materials and guide the exploration of exotic topological phases. The Berry dipole, a three-dimensional band crossing point beyond the Chern class, hosts a dipolar Berry curvature field and gives rise to numerous nontrivial quantum geometric effects. It has been proposed that the Berry dipole exhibits oriented Landau levels, whose spectrum critically relies on the orientation of the applied magnetic field. However, experimental demonstration of this phenomenon has remained elusive. Here we experimentally demonstrate oriented Landau levels by carefully engineering an inhomogeneous acoustic lattice. We observe distinct Landau level spectra and different propagation properties when the orientation of the pseudomagnetic field is reversed. Notably, we discover a new type of helical zero modes whose existence critically depends on the magnetic field's orientation. Our work paves the way for studying band crossings beyond Chern-class crossing points, including Berry multipoles and even-dimensional monopoles. Furthermore, it offers new insight for exploring topological devices.

# Introduction

Band crossing plays an important role in condensed matter theory, serving as the origin for various topological semimetals and insulators[1,2]. A well-known example is Weyl point, a two-band linear crossing point in three-dimensional (3D) momentum spaces, described by the Weyl Hamiltonian $H_{WP} = \eta \sum_{i=1}^{3} k_i \sigma_i$, where $k_i$ and $\sigma_i$ are momenta and Pauli matrices, respectively, and $\eta = \pm 1$ represnts chirality[3]. From the perspective of band topology, the Weyl point can be regarded as a source or sink of Berry curvature field in the momentum space, characterized by integer Chern numbers[4–9]. The presence of Berry monopole-antimonopole pairs gives rise to diverse exotic topological phenomena, including Fermi arc surface states[6,9,10], chiral anomaly[11–16], and quantum Hall effects[17].

Due to the isotropic distribution of Berry curvature of Berry monopoles (Fig. 1a), the above topological phenomena are also isotropic. An instance of this isotropy can be seen when a magnetic field couples with a charge-1 Weyl point, resulting in discrete Landau levels independent of the orientation of the magnetic field[18–25] and a chiral zero mode unidirectionally propagating along the direction of the magnetic field[12-15,23], as shown in Fig. 1b and c, respectively. Furthermore, Berry monopoles have been successfully generalized into linear multi-band ($N > 2$) crossing points characterized by large Chern numbers[26–28], and ($2N + 1$)-dimensional crossing points ($N > 1$) characterized by higher-order Chern numbers[29–32].

Very recently, band crossing beyond the above paradigm was predicted and studied theoretically[33–35]. These band crossing points, referred to as "Berry dipoles", carry a dipolar Berry curvature field, resulting in zero Chern number but nonzero Berry dipole moment. The Berry dipole can give rise to exotic quantum geometric effects such as anomalous Hall conductivity and orbital magnetization. Furthermore, the Berry dipole is considered as a topological phase transition point from trivial insulators to 3D Hopf insulators[36–40], an emerging theoretical frame that goes beyond ten-fold way topological classification[41] and is related to delicate topology. Chiral symmetric Berry dipoles can also be considered as a 3D projection of 4D tensor monopoles, which was originally predicted in string theory and recently generalized to condensed matter systems[42–47]. Interestingly, it was recently proposed that oriented Landau levels are formed when a Berry dipole is subject to an external magnetic field[48]. However, the experimental observation of this phenomenon remains elusive.

In this work, we experimentally realize oriented Landau levels in Berry dipole semimetals constructed by engineering an inhomogeneous acoustic crystal to introduce pseudomagnetic fields[49–52]. We demonstrate dramatically different Landau level spectra when the pseudomagnetic field is parallel or antiparallel to the Berry dipole moment. Specifically, when the pseudomagnetic field is parallel to the Berry dipole moment, there exist a pair of helical zero modes propagating respectively along and opposite to the pseudomagnetic field. Remarkably, the helical zero modes vanish and Landau levels are gapped when the pseudomagnetic field is reversed. Our work offers a platform to study topological matter beyond Chern class, such as Berry multipoles, four-dimensional tensor monopoles.

## Continuum model

We start from a minimal three-band continuum model of Berry dipole, whose Hamiltonian can be written as

$$H_{BD} = q_x \lambda_4 + q_y \lambda_5 + q_z \lambda_7 = \begin{pmatrix} 0 & 0 & q_x - iq_y \\ 0 & 0 & -iq_z \\ q_x + iq_y & iq_z & 0 \end{pmatrix}, \quad (1)$$

where $\vec{q} = (q_x, q_y, q_z)$ is the 3D momentum vector deviated from the Berry dipole and $\lambda_i$ represents the $3 \times 3$ Gell-Mann matrices[53]. In Fig. 1d, we plot the dispersion structure and distribution of Berry curvature near the triply degenerated nodal point at $\vec{q} = 0$. The band structure exhibits an isotropic configuration consisting of a flat band and two linear crossing bands with Weyl-like dispersions. The distribution of Berry curvature near the nodal point is characterized by a dipolar field distribution of $\Omega = \kappa(\vec{d} \cdot \vec{q}) \vec{q}/(2|q|^4)$, where $\kappa = 1$ and $\vec{d} = (0,0,1)$ represent the strength and orientation of the Berry dipole moment [see details in Supplementary material (SM), section 1]. Therefore, this crossing point can be regarded as a Berry dipole with zero Chern number but nonzero Berry dipole moment pointing to the positive $k_z$ direction.

Next, we consider a strong magnetic field applied to the Berry dipole system. Considering the axial rotation symmetry of the Berry curvature distribution, without loss of generality, the magnetic field can be written as

$$\vec{B} = B_0(\sin\theta, 0, \cos\theta), \qquad (2)$$

where $B_0$ is the strength of the magnetic field and $\theta$ is the angle between the magnetic field $\vec{B}$ and Berry dipole moment $\vec{d}$. The Landau levels can be exactly solved as

$$E = \pm\sqrt{2B_0\left(n + \frac{1-\cos\theta}{2}\right) + q_0^2}, \qquad (3)$$

where $q_0 = \vec{B}\cdot\vec{q}/|\vec{B}|$ is the projection of the momentum along the direction of pseudomagnetic field, and $n \in \{0, 1, 2, ...\}$ [SM, section 2 shows the comparison between Landau levels in Berry monopole and dipole systems]. The oriented Landau level spectrum against $\theta$ is illustrated in Fig. 1e with $q_0 = 0$. As we can see, different from the circular shape of Landau levels in Berry monopole system (Fig. 1b), the zeroth Landau levels of the Berry dipole exhibit a heart-shaped contour with a cusp at $\theta = 0$. Furthermore, we focus on two extreme cases ($\theta = 0$ and $\pi$) to reveal the oriention-dependent property. When $\theta = 0$, a pair of gapless helical zero modes (red and blue lines) connect bottom and top bulk bands with the group velocities in both positive and negative $z$ directions, as shown in the top pannel of Fig. 1f. In contrast, when the pseudomagnetic field is reversed ($\theta = \pi$), helical zero modes vanish and a band gap ($2\sqrt{2B_0}$) appears between two zeroth Landau levels, as shown in the bottom pannel of Fig. 1f. This behavior is akin to a field-effect transistor, where the ON and OFF states are controlled by the orientation of the applied magnetic field. This phenomeon significantly differs from the chiral zero mode observed in Weyl semimetals (Fig. 1c). It is worth noting that the oriented Landau level effect is a pure quantum geometric effect, which cannot be revealed by the semiclassical quantization method [see details in SM, section 3].

## Lattice model

To realize the Berry dipole and observe the associated oriented Landau levels, we design a 3D lattice model and introduce a pseudomagnetic field by engineering the inhomogeneous couplings. As shown in Fig. 2a, the unit cell of the hexagonal lattice contains three sublattices (blue, green, and red balls). The black (gray) and yellow bonds denote nearest-neighbor couplings $\pm t_1$ and $t_2$, and $a$ is the nearest-neighbor distance between sublattices. The detailed Hamiltonian is presented in the section 4 of the SM. In the case of $t_2 = t_1$, triply degenerated crossing points exist at K $(4\pi/(3\sqrt{3}a), 0, 0)$ and K' $(-4\pi/(3\sqrt{3}a), 0, 0)$ valleys, whose Berry dipole moments are in the positive and negative $k_z$ directions, respectively [see more details in SM, section 4]. The distribution of Berry dipoles in the 3D Brillouin zone and the global energy spectrum are present in Fig. 2b and c, respectively. When $t_2$ is slightly changed $((t_2 - t_1)/t_1 \ll 1)$, the effective Hamiltonian near the K (K') valley can be written as

$$H_{eff} = t_1 \left[ \left( \eta \frac{3a}{2} \delta k_x + \frac{t_2 - t_1}{t_1} \right) \lambda_4 + \frac{3a}{2} \delta k_y \lambda_5 + 2a \delta k_z \lambda_7 \right], \tag{4}$$

where $\eta = \pm 1$ for the K (K') valley. Interestingly, the relative coupling difference $(t_2 - t_1)/t_1$ causes a linear shift of Berry dipoles in the $k_x$ direction.

To realize a pseudomagnetic field in $z$ direction, we introduce spatially inhomogeneous couplings $t_2$ along the $y$ direction, as shown in Fig. 2d. In the first case, the crystal is designed such that as $y$ increases, the Berry dipoles shift away from the $\Gamma$ point, which introduces an artificial gauge potential of $\vec{A} = (\mp B_0 y, 0, 0)$ in the K (K') valley, corresponding to a pseudomagnetic field in the positive (negative) $z$ direction with the strength of $0.0033 a^{-2}$. The generated pseudomagnetic field is parallel to the Berry dipoles at both K and K' valleys, corresponding to the case of $\theta = 0$. In the direction perpendicular to the pseudomagnetic field (the $k_x$ direction), Landau levels are discrete flat energy levels with $k_z = 0$, as shown by the red line (doubly degenerated zeroth modes at zero energy) and blue lines (higher-order modes) in Fig. 2e. Note that due to the large control range of the coupling coefficient $t_2$, the pseudomagnetic field is not perfectly uniform, which leads to a slight slope for the higher-order Landau levels. In Fig. 2f, we depict Landau level spectrum in the $k_z$ direction with $k_x = \pm 4\pi/(3\sqrt{3}a)$, and red lines represent the helical zero modes.

In the second case, we spatially control the inhomogeneous coupling $t_2$ as a decreasing function of position $y$, as shown in Fig. 2g. As the $y$ position increases, Berry dipoles move toward the Γ point, which introduces an artificial gauge potential of $\vec{A} = (\pm B_0 y, 0,0)$ in the K (K′) valley, corresponding to a pseudomagnetic field in the negative (positive) z direction with the strength of $0.0033a^{-2}$. At both K and K′ valleys, the pseudomagnetic field is antiparallel to Berry dipoles (the case of $\theta = \pi$), resulting in a gapped Landau level spectrum. Fig. 2h and i respectively show the energy spectra of Landau levels in the $k_x$ direction with $k_z = 0$ and in the $k_z$ direction with $k_x = -4\pi/(3\sqrt{3}a)$. A gap occurs at the K and K′ valleys, and no propagation states are allowed in the band gap. Moreover, one may also realize pseudomagnetic fields in other directions, such as $\vec{B} = (0, B_0, 0)$ [see more details in SM, section 5].

Acoustic crystal design

Next, we design a 3D acoustic crystal to realize the above lattice model. As shown in Fig. 3a, the acoustic unit cell consists of three acoustic resonators (blue blocks), corresponding to three sublattices "A", "B" and "C" in Fig. 2a, and the channels between them (gray and yellow blocks) provide the nearest-neighbor couplings. Acoustic units are staggered in the $x$ direction and stacked in the $z$ direction. This configuration can realize the Hamiltonian of the hexagonal lattice in an easily processable tetragonal-like lattice [see more details in SM, section 6]. $a_1$ ($a_2$) is the distance between A and C (B and C) resonators. The resonators are filled with air and surrounded by resin walls, which can be considered as acoustic hard boundaries. Here, we consider the dipole modes in acoustic resonators, whose simulated band structure matches well with that of the tight-binding model [shown in SM, Fig. S3]. We define a dimensionless parameter $\xi$ to describe the area of cross-section of yellow channels $S = \xi S_0$, which has an approximately linear correlation with the strength of intra-cell coupling between A and C resonators but does not affect the frequency of Berry dipoles [see details in SM, section 7].

To experimentally demonstrate the oriented Landau levels, we construct a supercell containing 16 units in the $y$ direction and spatially modulate $\xi$ in the range of 0.25 to 1.75, which introduces a strong pseudomagnetic field in the $z$ direction with the strength of $|\vec{B}| = 0.067 a_1^{-2}$. As illustrated in Fig. 3b, when $y$ increases, $\xi$ is linearly increased (decreased) to introduce a pseudomagnetic field parallel (antiparallel) to the Berry dipoles, corresponding to the case of $\theta = 0$ $(\pi)$ [see more detailed structure parameters in SM, Table S1 and 2]. In addition, one can also construct acoustic supercell in the $z$ direction with inhomogeneous $\xi$ to realize the pseudomagnetic perpendicular to Berry dipoles, corresponding to the case of $\theta = \pi/2$ [see more details in SM, section 8].

## Experimental results

To experimentally observe the oriented Landau levels, we fabricated two samples containing $18 \times 16 \times 12$ acoustic units, as shown in Fig. 3c. The two samples have the same configurations except for the different variations of inhomogeneous channels in the $y$ direction as shown in Fig. 3b, corresponding to the cases of $\theta = 0$ and $\pi$, respectively. In the case of $\theta = 0$, in order to distinguish the zeroth Landau levels from the middle flat bulk band, only C sublattices are excited and measured, whose sound pressure field has no contribution from the middle band [see the proof in SM, section 9]. Using acoustic pump-probe measurements, we obtain the 2D projected dispersion spectra in the $k_x - k_z$ plane [see more details about experimental setup and measurements in SM, section 10]. The colored plots in Fig. 3d and f show the measured dispersion spectra in the $k_x$ direction with fixed $k_z = 0$ for Samples 1 and 2, respectively, together with the numerical dispersion spectra (gray dots). Zoom-in views of the dispersion near K′ and K valleys (insets i and ii in Fig. 3d and f) show the presence of gapless flat zeroth Landau levels around the frequency of 8600 Hz in Sample 1, while a band gap between two zeroth modes in Sample 2.

For more quantitative characterization of the Landau levels, we further present the spectra of Samples 1 and 2 at the K valley in Fig. 3e and g, respectively. Each spectrum exhibits several peaks corresponding to the locations of the Landau levels. For Sample 1, there is a pronounced peak right at the middle of the spectrum, corresponding to the zeroth modes. In contrast, in Sample 2 the two zeroth modes are separated forming a gap in the middle. The locations of the Landau levels in the measured spectra agree well with those of the simulation. Thus, our measurement results on the cases with $\theta = 0$ and $\pi$ provide the direct evidence of the oriented Landau levels.

We next focus on the oriented helical zero modes and their propagating properties in the direction of pseudomagnetic field in Samples 1 and 2. For Sample 1, a pair of gapless helical zero modes are present in each valley with negative and positive group velocities, as shown in Fig. 4a and b, respectively. Again, a good agreement is reached between the simulation and measurement. The sound pressure field distributions in the $y$-$z$ plane are also measured at the exciting frequency of 8500 Hz, as shown in left panels of Fig. 4c and d, corresponding to sound sources placed in the top and bottom surfaces, respectively. It is seen that the sound wave propagates along the $z$ direction and follows a Gaussian distribution in the $y$ direction, agreeing well with the simulated sound pressure field distributions in right panels of Fig. 4c and d.

The measured dispersion for Sample 2 is shown in Fig. 4e and f. Because of the gap between the two Landau zeroth modes, the propagation of sound within this frequency range is prohibited. Under the excitation at 8500 Hz, the measured sound pressure is localized at the source position and exponentially decays into the bulk, as shown in the left panels of Fig. 4g and h. The measurement results agree well with the simulated sound pressure field distributions in right panels of Fig. 4g and h. In the above measurements, the excitations and measurements are taken at C sublattices. Sound pressure field distributions at other sublattices are presented in the section 11 of the SM.

**Discussion**

To sum up, we have experimentally demonstrated oriented Landau levels in the 3D Berry dipole system via introduction of pseudomagnetic fields. We have designed and fabricated two samples with opposite pseudomagnetic fields that are parallel ($\theta =$

0) and antiparallel ($\theta = \pi$) to Berry dipoles, respectively, and observed dramatically different Landau level spectra, providing direct evidence for the oriented Landau level effect. Furthermore, we observe a new type of helical zero modes existing when $\theta = 0$ and vanishing when $\theta = \pi$, which critically depend on the magnetic field's orientation. Our work opens the door to exploring the effect of gauge field on band crossing points beyond conventional Chern-class crossing points. Meanwhile, our acoustic design provides a platform for studying exotic topological phenomena related to Berry dipoles, such as 3D Hopf insulators and even tensor monopoles in synthetic 4D acoustic lattice. The distinct Landau levels exhibited under different orientations of pseudomagnetic field may offer novel insights for the design of optical devices.

## Methods

### Structure parameters of acoustic lattice

The size of A and C resonators is $w \times w \times h_1 = 8\ mm \times 8\ mm \times 20 mm$, while the height of B resonator is sightly different $h_2 = 20.7\ mm$. The size of cross section of gray coupling channels connecting A and C (B and C) resonators is $d_1 \times d_2 = 4.5\ mm \times 5\ mm$ ($d_2 \times d_2 = 5\ mm \times 5\ mm$). The size of cross section of yellow coupling channels is controlled by the dimensionless parameter $\xi$. The width and height of the cross section are $d_1 = \sqrt{\xi} 4.5\ mm$ and $d_2 = \sqrt{\xi} 5\ mm$, respectively. The nearest-neighbor distance between A and C (B and C) resonators is $a_1 = 17.6\ mm$ ($a_2 = 22.2\ mm$). The size of the whole acoustic samples is $328.8\ mm \times 297.5\ mm \times 556.8\ mm$.

### Numerical simulations

All simulations in this work were performed using the finite element method, i.e. a commercial software COMSOL Multiphysics (pressure acoustic module). Because of the huge acoustic impedance mismatch between the photosensitive resin and air, the resin material can be considered as a hard boundary for the simulations. To simulate dispersion spectra of Landau levels, periodic boundary conditions were imposed in both the $x$ and $z$ directions and the open boundary condition was applied to the $y$ direction, which contains 16 units with inhomogeneous coupling channels. To simulate the distributions of the sound pressure at the frequency of 8500 Hz in Samples 1 and 2, the periodic boundary condition was imposed in the $x$ direction and open boundary conditions were applied to the $y$ and $z$ direction, which contains 16 × 12 units with inhomogeneous coupling channels. Acoustic lattices were filled with air (density = 1.22 kg m$^{-3}$, speed of sound = 342 m s$^{-1}$ at room temperature). When we calculate sound response in Fig. 3e, g and Fig. 4 c, d, g, h, an additional loss is added to the speed of sound ($c = 342 + 1.5i\ ms^{-1}$) to mimic sound loss in the real experiment.

## Acknowledgements

S.Z. acknowledges New Cornerstone Science Foundation, the Research Grants Council of Hong Kong (AoE/P-502/20 and 17309021). C.L. acknowledges the National Natural Science Foundation of China (12274031), Beijing Institute of Technology Research Fund Program for Teli Young Fellows, and Beijing Institute of Technology Innovative Talents Science and Technology Funding Special Plan (2022CX01006). X.H. acknowledges National Key R&D Program of China (No. 2022YFA1404500), Guangdong Basic and Applied Basic Research Foundation (No. 2021B1515020086).

## References

1. Hasan, M. Z. & Kane, C. L. Colloquium: Topological insulators. *Rev Mod Phys* **82**, 3045–3067 (2010).
2. Qi, X.-L. & Zhang, S.-C. Topological insulators and superconductors. *Rev Mod Phys* **83**, 1057–1110 (2011).
3. Herring, C. Accidental Degeneracy in the Energy Bands of Crystals. *Physical Review* **52**, 365–373 (1937).
4. Lu, L., Fu, L., Joannopoulos, J. D. & Soljačić, M. Weyl points and line nodes in gyroid photonic crystals. *Nat Photonics* **7**, 294–299 (2013).
5. Xu, S.-Y. *et al.* Discovery of a Weyl fermion semimetal and topological Fermi arcs. *Science* **349**, 613–617 (2015).
6. Lu, L. *et al.* Experimental observation of Weyl points. *Science* **349**, 622–624 (2015).
7. Lv, B. Q. *et al.* Experimental Discovery of Weyl Semimetal TaAs. *Phys Rev X* **5**, 31013 (2015).
8. Burkov, A. A. Topological semimetals. *Nat Mater* **15**, 1145–1148 (2016).
9. Yang, B. *et al.* Ideal Weyl points and helicoid surface states in artificial photonic crystal structures. *Science* **359**, 1013–1016 (2018).

10. Xu, S.-Y. *et al.* Observation of Fermi arc surface states in a topological metal. *Science* **347**, 294–298 (2015).
11. Li, Q. *et al.* Chiral magnetic effect in ZrTe5. *Nat Phys* **12**, 550–554 (2016).
12. Gooth, J. *et al.* Experimental signatures of the mixed axial–gravitational anomaly in the Weyl semimetal NbP. *Nature* **547**, 324–327 (2017).
13. Jia, H. *et al.* Observation of chiral zero mode in inhomogeneous three-dimensional Weyl metamaterials. *Science* **363**, 148–151 (2019).
14. Peri, V., Serra-Garcia, M., Ilan, R. & Huber, S. D. Axial-field-induced chiral channels in an acoustic Weyl system. *Nat Phys* **15**, 357–361 (2019).
15. Zhao, Y. X. & Yang, S. A. Index Theorem on Chiral Landau Bands for Topological Fermions. *Phys Rev Lett* **126**, 46401 (2021).
16. Yan, M. *et al.* Antichirality Emergent in Type-II Weyl Phononic Crystals. *Phys Rev Lett* **130**, 266304 (2023).
17. Yang, K.-Y., Lu, Y.-M. & Ran, Y. Quantum Hall effects in a Weyl semimetal: Possible application in pyrochlore iridates. *Phys Rev B* **84**, 75129 (2011).
18. McCann, E. & Fal'ko, V. I. Landau-Level Degeneracy and Quantum Hall Effect in a Graphite Bilayer. *Phys Rev Lett* **96**, 86805 (2006).
19. Schine, N., Ryou, A., Gromov, A., Sommer, A. & Simon, J. Synthetic Landau levels for photons. *Nature* **534**, 671–675 (2016).
20. Yang, Z., Gao, F., Yang, Y. & Zhang, B. Strain-Induced Gauge Field and Landau Levels in Acoustic Structures. *Phys Rev Lett* **118**, 194301 (2017).
21. Wen, X. *et al.* Acoustic Landau quantization and quantum-Hall-like edge states. *Nat Phys* **15**, 352–356 (2019).
22. Jamadi, O. *et al.* Direct observation of photonic Landau levels and helical edge states in strained honeycomb lattices. *Light Sci Appl* **9**, 144 (2020).
23. Ma, S. *et al.* Gauge Field Induced Chiral Zero Mode in Five-Dimensional Yang Monopole Metamaterials. *Phys Rev Lett* **130**, 243801 (2023).
24. Barsukova, M. *et al.* Direct observation of Landau levels in silicon photonic crystals. *Nat Photonics* (2024) doi:10.1038/s41566-024-01425-y.
25. Barczyk, R., Kuipers, L. & Verhagen, E. Observation of Landau levels and chiral edge states in photonic crystals through pseudomagnetic fields induced by synthetic strain. *Nat Photonics* (2024) doi:10.1038/s41566-024-01412-3.
26. Bradlyn, B. *et al.* Beyond Dirac and Weyl fermions: Unconventional quasiparticles in conventional crystals. *Science* **353**, aaf5037 (2016).
27. Ezawa, M. Chiral anomaly enhancement and photoirradiation effects in multiband touching fermion systems. *Phys Rev B* **95**, 205201 (2017).
28. Rao, Z. *et al.* Observation of unconventional chiral fermions with long Fermi arcs in CoSi. *Nature* **567**, 496–499 (2019).
29. Lian, B. & Zhang, S.-C. Five-dimensional generalization of the topological Weyl semimetal. *Phys Rev B* **94**, 41105 (2016).
30. Lian, B. & Zhang, S.-C. Weyl semimetal and topological phase transition in five dimensions. *Phys Rev B* **95**, 235106 (2017).
31. Ma, S. *et al.* Linked Weyl surfaces and Weyl arcs in photonic metamaterials. *Science* **373**, 572–576 (2021).

32. Sugawa, S., Salces-Carcoba, F., Perry, A. R., Yue, Y. & Spielman, I. B. Second Chern number of a quantum-simulated non-Abelian Yang monopole. *Science* **360**, 1429–1434 (2018).
33. Nelson, A., Neupert, T., Bzdušek, T. & Alexandradinata, A. Multicellularity of Delicate Topological Insulators. *Phys Rev Lett* **126**, 216404 (2021).
34. Nelson, A., Neupert, T., Alexandradinata, A. & Bzdušek, T. Delicate topology protected by rotation symmetry: Crystalline Hopf insulators and beyond. *Phys Rev B* **106**, 75124 (2022).
35. Zhu, P., Noh, J., Liu, Y. & Hughes, T. L. Scattering theory of delicate topological insulators. *Phys Rev B* **107**, 195110 (2023).
36. Deng, D.-L., Wang, S.-T., Shen, C. & Duan, L.-M. Hopf insulators and their topologically protected surface states. *Phys Rev B* **88**, 201105 (2013).
37. Liu, C., Vafa, F. & Xu, C. Symmetry-protected topological Hopf insulator and its generalizations. *Phys Rev B* **95**, 161116 (2017).
38. Schuster, T., Gazit, S., Moore, J. E. & Yao, N. Y. Floquet Hopf Insulators. *Phys Rev Lett* **123**, 266803 (2019).
39. Schuster, T. *et al.* Realizing Hopf Insulators in Dipolar Spin Systems. *Phys Rev Lett* **127**, 15301 (2021).
40. Lim, H., Kim, S. & Yang, B.-J. Real Hopf insulator. *Phys Rev B* **108**, 125101 (2023).
41. Ryu, S., Schnyder, A. P., Furusaki, A. & Ludwig, A. W. W. Topological insulators and superconductors: tenfold way and dimensional hierarchy. *New J Phys* **12**, 065010 (2010).
42. Kalb, M. & Ramond, P. Classical direct interstring action. *Physical Review D* **9**, 2273–2284 (1974).
43. Nepomechie, R. I. Magnetic monopoles from antisymmetric tensor gauge fields. *Physical Review D* **31**, 1921–1924 (1985).
44. Palumbo, G. & Goldman, N. Revealing Tensor Monopoles through Quantum-Metric Measurements. *Phys Rev Lett* **121**, 170401 (2018).
45. Palumbo, G. Non-Abelian Tensor Berry Connections in Multiband Topological Systems. *Phys Rev Lett* **126**, 246801 (2021).
46. Tan, X. *et al.* Experimental Observation of Tensor Monopoles with a Superconducting Qudit. *Phys Rev Lett* **126**, 17702 (2021).
47. Chen, M. *et al.* A synthetic monopole source of Kalb-Ramond field in diamond. *Science* **375**, 1017–1020 (2022).
48. Graf, A. & Piéchon, F. Massless multifold Hopf semimetals. *Phys Rev B* **108**, 115105 (2023).
49. Pikulin, D. I., Chen, A. & Franz, M. Chiral Anomaly from Strain-Induced Gauge Fields in Dirac and Weyl Semimetals. *Phys Rev X* **6**, 41021 (2016).
50. Grushin, A. G., Venderbos, J. W. F., Vishwanath, A. & Ilan, R. Inhomogeneous Weyl and Dirac Semimetals: Transport in Axial Magnetic Fields and Fermi Arc Surface States from Pseudo-Landau Levels. *Phys Rev X* **6**, 41046 (2016).
51. Weström, A. & Ojanen, T. Designer Curved-Space Geometry for Relativistic Fermions in Weyl Metamaterials. *Phys Rev X* **7**, 41026 (2017).
52. Jia, H., Zhang, R.-Y., Gao, W., Zhang, S. & Chan, C. T. Chiral transport of pseudospinors induced by synthetic gravitational field in photonic Weyl metamaterials. *Phys Rev B* **104**, 45132 (2021).
53. Gell-Mann, M. Symmetries of Baryons and Mesons. *Physical Review* **125**, 1067–1084

(1962).

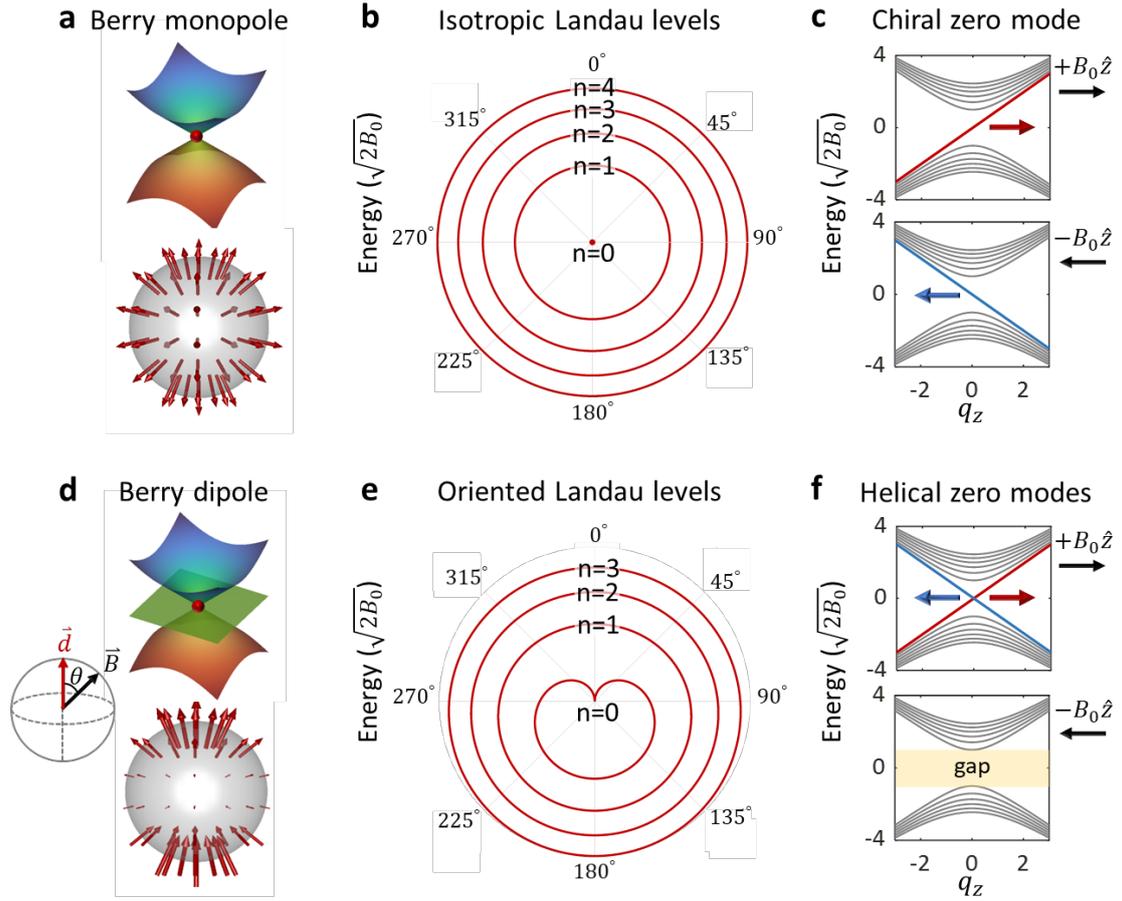

**Fig. 1: Landau levels in Berry monopole and dipole systems. a, d** The dispersion spectra (top panels) and Berry curvature distribution (bottom panels) in the Berry monopole (**a**) and dipole (**d**) systems, respectively. The inset in **d** shows the magnetic field $\vec{B}$, the Berry dipole $\vec{d}$, and $\theta$ is the angle between them. **b, e** The Landau level spectra of the Berry monopole (**b**) and dipole (**e**) systems against the polar angle of the magnetic field. **c, f** The dispersion spectra of the Berry monopole (**c**) and dipole (**f**) systems in the $q_z$ direction when the magnetic field is along (top panels) and opposite (bottom panels) to the $z$ direction. Red and blue lines in **c** and **f** denote chiral zero modes and helical zero modes, respectively.

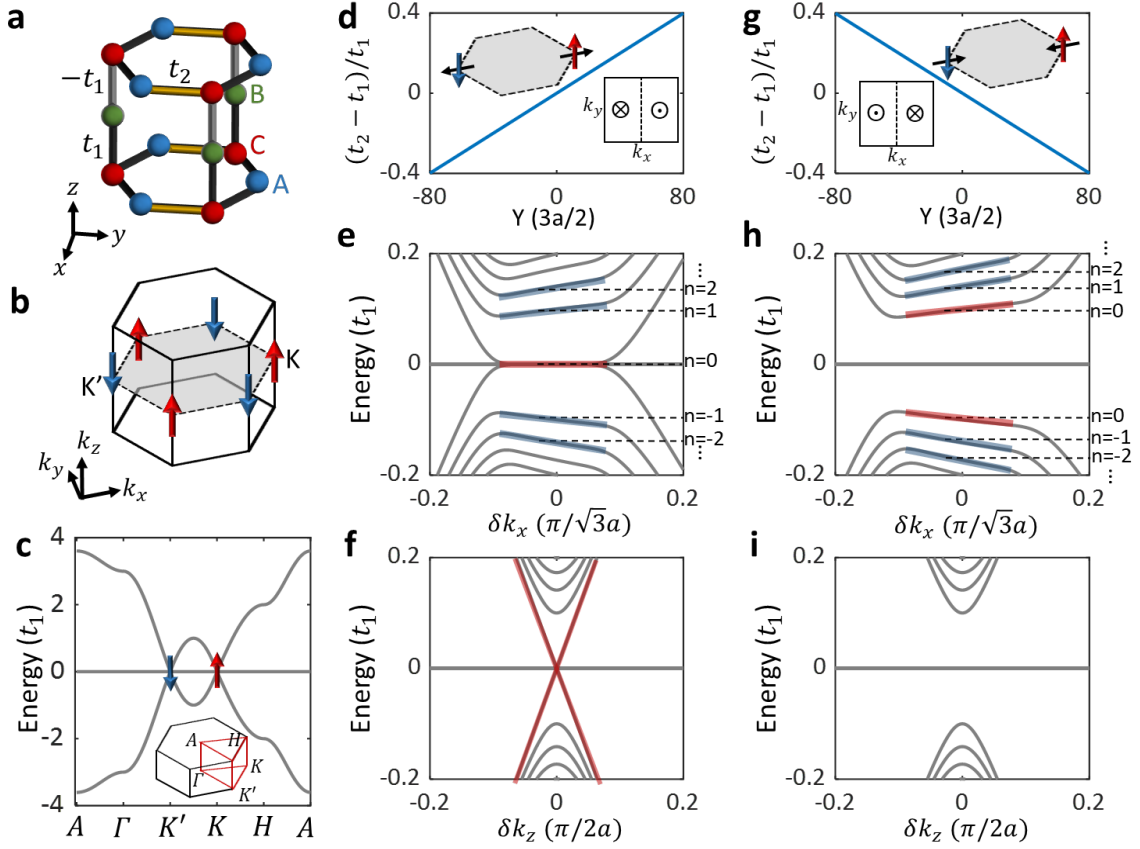

**Fig. 2: Oriented Landau levels in Berry dipole semimetals induced by inhomogeneous couplings. a** The tight-binding model for the Berry dipole semimetal. Blue, green and red dots represent A, B and C sublattices. The black (gray) and yellow tubes represent the strength of nearest-neighbor coupling $\pm t_1$ and $t_2$. The nearest-neighbor distance between sublattices is $a$. **b, c** The 3D Brillouin zone and global energy spectrum with $t_1 = t_2$. Two Berry dipoles pointing in the positive and negative $z$ directions (red and blue arrows) exist at $\vec{k} = (\pm 4\pi/(3\sqrt{3}a), 0, 0)$ (K and K′ valleys). **d, g** The relation of the coupling $t_2$ to the $y$ position, which is a increasing and decreasing function in **d** and **g**, respectively. The insets show the trajectory of Berry dipoles as the $y$ position increases and corresponding pseudomagnetic fields. The angle between pseudomagnetic field $\vec{B}$ and Berry dipole $\vec{d}$ is $\theta = 0$ in **d** and $\theta = \pi$ in **g**. **e, h** Energy spectra in the $k_x$ direction with $k_z = 0$, corresponding to cases of $\theta = 0$ and $\pi$, respectively. Red and blue lines represent zeroth and higher-order Landau levels. **f, i** The energy spectrum in the $k_z$ direction with $k_x = -4\pi/(3\sqrt{3}a)$, corresponding to cases of $\theta = 0$ and $\pi$, respectively. Red lines indicate helical zero modes in the case of

$\theta = 0.$

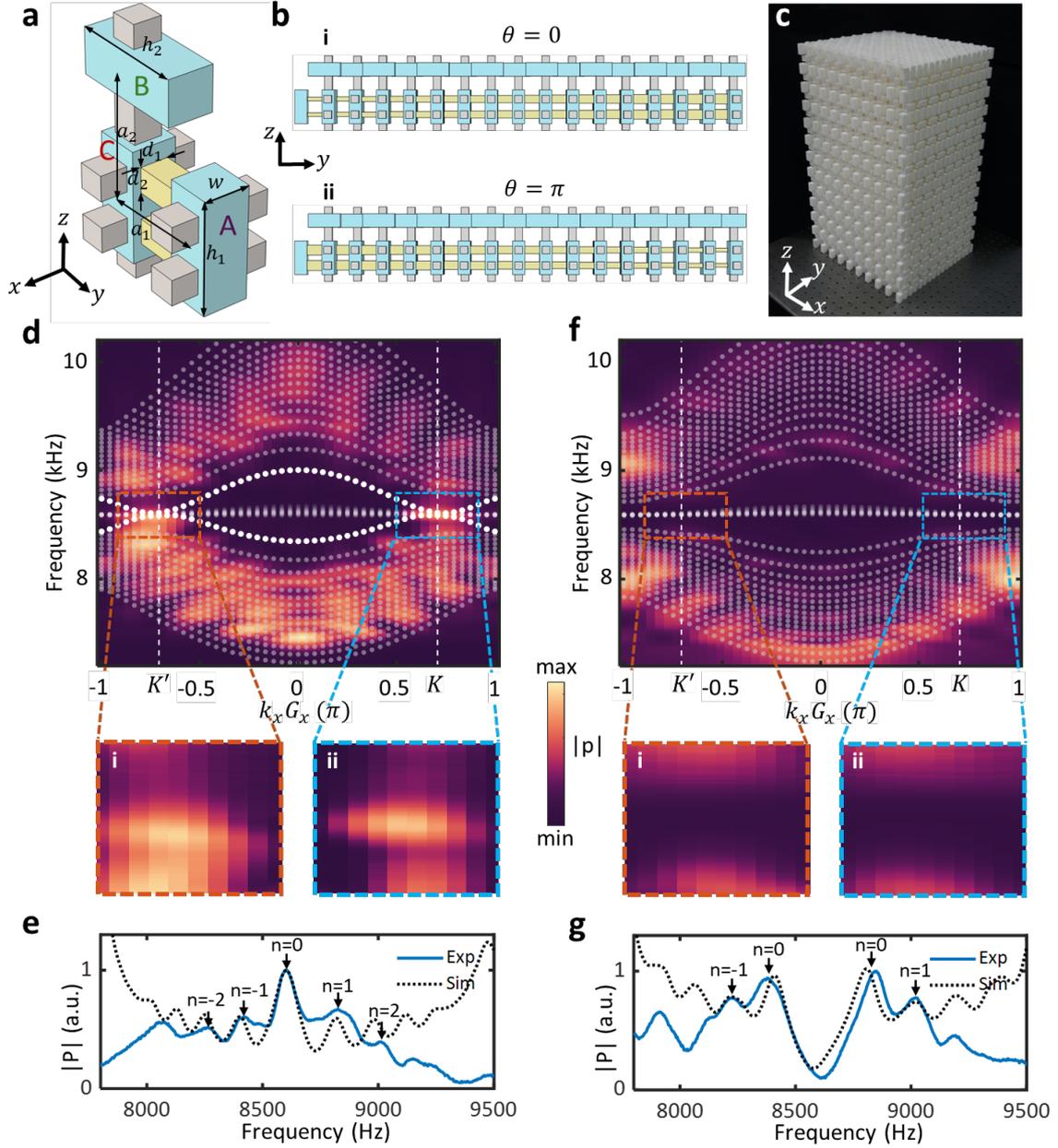

**Fig. 3: Acoustic design and experimental detection of oriented Landau levels. a** unit cell of the acoustic lattice. Blue acoustic resonators (blue blocks) denote sublattices (A, B and C), and acoustic coupling channels (yellow and gray blocks) denote the nearest-neighbor couplings between sublattices. The area of cross-section of yellow channels is $S_0$, indicating the strength of coupling $t_2$. $a_1$ ($a_2$) is the nearest-neighbor distance between A and C (B and C) sublattices. **b** Two configurations of inhomogeneous yellow channels in the acoustic supercell, corresponding to cases of $\theta = 0$ and $\pi$, respectively. The strength of pseudomagnetic field is $0.067 a_1^{-2}$. **c** Photographs of two experimental samples with the same appearance but different

configurations in **b**. Samples 1 and 2 correspond to the case of $\theta = 0$ and $\pi$. **d, f** Measured Landau levels in the $k_x$ direction with $k_z = 0$ corresponding to Samples 1 and 2, respectively. The gray and white dots are simulated dispersion spectra and gapless zeroth Landau levels in the case of $\theta = 0$. The lattice constant in the $k_x$ direction is $G_x = 2a_1$. The insets i and ii are zoom-in views near the K' and K valleys. **e, g** Measured (blue solid lines) and simulated (black dashed lines) acoustic pressure spectra at the K valley for Samples 1 and 2, respectively.

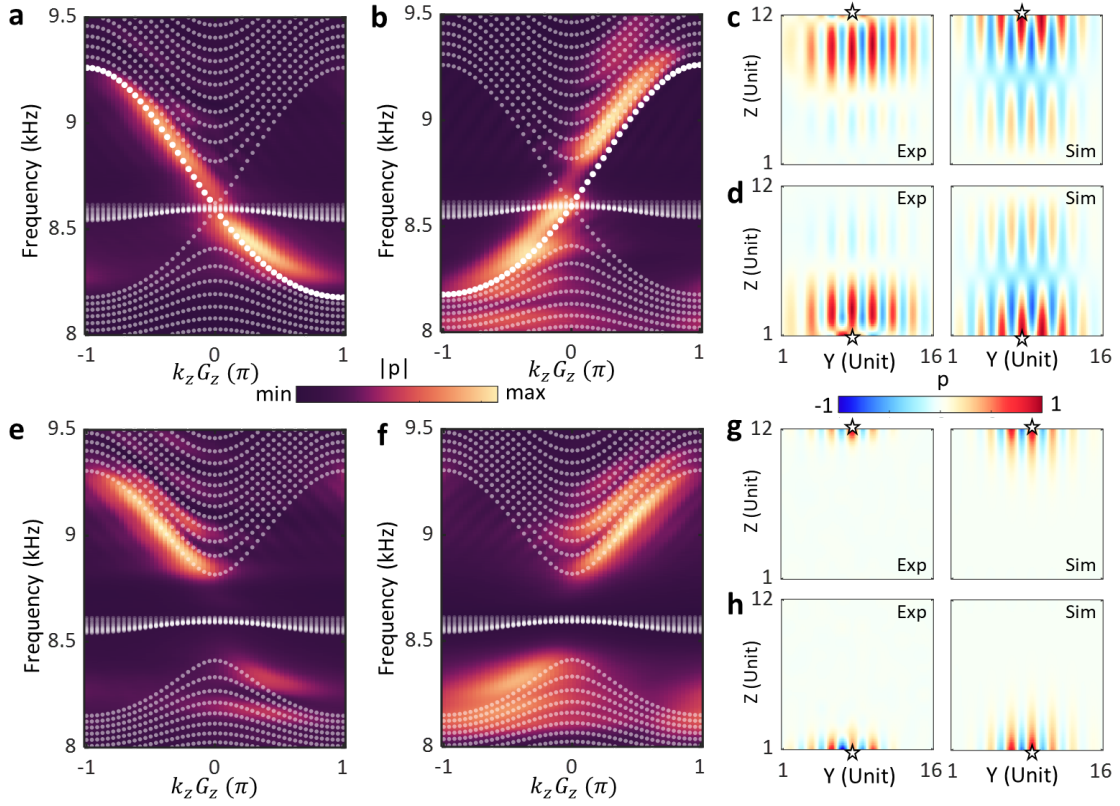

**Fig. 4: Experimental detection of oriented helical zero modes and propagating properties. a, b** Measured dispersion spectra of helical zero modes in the K valley in Sample 1. The down-going and up-going zeroth modes are excited, respectively. The white dots denote the excited zeroth modes in simulation, and the gray dots are simulated dispersion spectra in the case of $\theta = 0$. The lattice constant in the $k_z$ direction is $G_z = 2a_2$. **c, d** Sound pressure distributions of down-going and up-going zeroth modes at the frequency of 8500 Hz, respectively. Left and right panels represent experimental and simulated results. Pentagrams are sound sources. **e, f** Measured dispersion spectra of gapped Landau levels in the K valley in Sample 2. The down-going and up-going parts of Landau levels are excited, respectively. The gray dots are simulated dispersion spectra in the case of $\theta = \pi$. **g, h** Sound pressure distributions at the frequency of 8500 Hz, in experiment (left panels) and simulation (right panels). The sources are located at the top and bottom surfaces, respectively.